\documentclass[conference]{IEEEtran}
\usepackage{cite}
\usepackage{gensymb}
\usepackage{amsmath,amssymb,amsfonts}
\usepackage{algorithmic}
\usepackage{graphicx}
\usepackage{textcomp}
\usepackage{xcolor}
\usepackage{hyperref}
\usepackage{multirow}
\usepackage[absolute,showboxes]{textpos}

\TPMargin{5pt}

\newcommand{\copyrightstatement}{
    \begin{textblock}{0.84}(0.08,0.93)    % tweak here: {box width}(leftposition, rightposition)
         \noindent
         \footnotesize
         \copyright  2024 IEEE.  Personal use of this material is permitted.  Permission from IEEE must be obtained for all other uses, in any current or future media, including reprinting/republishing this material for advertising or promotional purposes, creating new collective works, for resale or redistribution to servers or lists, or reuse of any copyrighted component of this work in other works. The published paper is available at \url{https://doi.org/10.1109/EPECS62845.2024.10805508}.
    \end{textblock}
}

\begin{document}
\copyrightstatement
\bstctlcite{IEEEexample:BSTcontrol}
\title{Seasonal Performance Evaluation of a Hybrid PV-Wind-Battery Power System for a Mars Base}

\author{\IEEEauthorblockN{Abdollah Masoud Darya\IEEEauthorrefmark{2}\IEEEauthorrefmark{1},
Ramesh~C.~Bansal\IEEEauthorrefmark{3}\IEEEauthorrefmark{4}, and 
Omaima Anwar Jarndal\IEEEauthorrefmark{3}}
\IEEEauthorblockA{
\IEEEauthorrefmark{2}SAASST, University of Sharjah, Sharjah, UAE\\
\IEEEauthorrefmark{3}Dept. of EE, University of Sharjah, Sharjah, UAE\\
\IEEEauthorrefmark{4}Dept. of EECE, University of Pretoria, Pretoria, South Africa\\
\IEEEauthorrefmark{1}Corresponding Author: adarya@sharjah.ac.ae}}

\IEEEoverridecommandlockouts
\IEEEpubid{\makebox[\columnwidth]{\centering\begin{tabular}[t]{@{}l@{}}EPECS 2024 Sharjah, UAE\\979-8-3315-1696-3/24/\$31.00~\copyright2024 IEEE\end{tabular}\hfill}\hspace{\columnsep}\makebox[\columnwidth]{ }}
%979-8-3315-1696-3/24/$31.00 ©2024 IEEE
\maketitle

\begin{abstract}
This work investigates a hybrid photovoltaic-wind-battery power system designed to sustain a Mars base under varying seasonal and climatic conditions. The Mars Climate Database was utilized to simulate the effects of seasonal changes, diurnal cycles, and dust storms on the system’s power generation. The seasonal performance was analyzed across the Martian surface and at potential habitation sites proposed in the ``First Landing Site/Exploration Zone Workshop for Human Missions to the Surface of Mars (FLSW).'' Within the hybrid system, the photovoltaic arrays serve as the primary energy source, with wind turbines providing essential backup during nighttime and dust storms. A single $1\,000\,\mathrm{m}^2$ photovoltaic array, a $33.4\,\mathrm{m}$ diameter wind turbine, and a $312\,\mathrm{kWh}$ battery can support a six-person Mars base at $32.1\%$ of the Martian surface during the equinoxes and solstices, expanding to $51.7\%$ with three sets of arrays and turbines. Additionally, $24$ FLSW sites can be supported throughout the solstices and equinoxes by a single photovoltaic array, turbine, and battery, even during global dust storms. Among the $24$ sites, Hebrus Valles, Huygens Crater, and Noctis Labyrinthus had the highest energy production potential. These findings are expected to guide further research on hybrid renewable power systems for Mars exploration.
\end{abstract}


%\IEEEpeerreviewmaketitle

\section{Introduction}
Future crewed missions to Mars will rely on in-situ resource utilization for energy production. Given the scarcity of natural resources on Mars, both nuclear and renewable energy sources are among the few viable options \cite{Imre2024engineering}. While nuclear power has been used for Martian robotic missions, it poses risks when located near human settlements and presents challenges for long-term nuclear waste disposal. The two main renewable energy sources under consideration are photovoltaic (PV) and wind power systems. Complete reliance on PV sources is impractical due to the day-night cycle and the frequent dust storms on the Martian surface. These dust storms limit the solar flux reaching the Martian surface and cause dust accumulation on the panels. Fortunately, during these periods, wind energy can serve as a suitable backup source \cite{hartwick2023assessment}. Additionally, a battery energy storage system (BESS) can be incorporated for added flexibility.\par

\subsection{Literature Review}
Abel \emph{et~al.} \cite{abel2022photovoltaics} evaluated the use of PV-based power systems to support a Mars base. They used climate data to study the feasibility of supporting a six-person crewed outpost over a significant portion of the Martian surface. However, they did not include wind-based power generation in their analysis.\par
Schorbach \emph{et~al.} \cite{schorbach2022wind} studied the feasibility of using wind energy as a backup to a solar power system on Mars. They found a low correlation between wind speed data from the Viking Lander 2 mission and optical depth. In other words, dust storms did not coincide with the highest wind speeds. They concluded that typical ground-based wind turbines are not suitable as a backup power supply during dust storms. However, they only evaluated three sites on the planet and recommended expanding the study to cover the entire surface using the Mars Climate Database (MCD). Similarly, Delgado-Bonal \emph{et~al.} \cite{delgado2016solar} recommended using solar energy as the main source of power on Mars, and advised against using wind power due to the low density of the Martian atmosphere and low wind speeds. It must be noted that the findings of these two studies were based on their analysis of a limited number of locations on the Martian surface.\par
Hartwick \emph{et~al.} \cite{hartwick2023assessment} conducted a surface-wide study on the feasibility of using wind energy as a standalone or backup power source to solar power using the NASA Ames Mars Global Climate Model. They proposed a power system that includes an Enercon E33 $330\,\mathrm{kW}$ wind turbine and a photovoltaic array with $2\,500\,\mathrm{m}^2$ total area. They found that wind energy can compensate for the seasonal and diurnal reductions in solar power, particularly during dust storms. However, they did not include a BESS in their analysis.\par

\subsection{Contributions}
This study builds on previous work \cite{hartwick2023assessment,schorbach2022wind} by evaluating the feasibility of using a hybrid power system comprising PV arrays, wind turbines, and a BESS to support a Mars base. The performance of different configurations, such as PV-wind, PV-BESS, and PV-wind-BESS are compared during the solstices, equinoxes, and global dust storms. The MCD is used to study the effects of different seasonal and climatic parameters on different configurations. Like \cite{hartwick2023assessment}, this study also covers the entire Martian surface. This work additionally extends the seasonal analysis to the habitation sites proposed in the ``First Landing Site/Exploration Zone Workshop for Human Missions to the Surface of Mars (FLSW).'' The most promising of these sites in terms of their energy production potential are highlighted.\par

\section{Background}\label{Sec2}
The following subsections introduce background information on key points relevant to this study.

\subsection{Base Energy Requirements}\label{RefEn}
A study by the National Aeronautics and Space Administration (NASA) estimated that the surface habitat power requirement for a six-person crew over $500$ Martian days ranges from $12.26$ to $22.77\,\mathrm{kW}$ \cite{rucker2015integrated}. Therefore, this study assumes a minimum uninterrupted power supply of $13\,\mathrm{kW}$ for the habitat.
\par

\subsection{Mars Climate Database}
To understand the impact of dust storms, seasonal, and diurnal effects on energy production, this study uses MCD v6.1 \cite{forget1999improved} to retrieve solar flux, wind speed, and air density data\footnote{The data used in this work are available from \\\url{https://github.com/AbdollahMasoud/EPECS-2024}.}. The MCD provides climate information derived from numerical simulations validated with observational data. It is publicly available online at \url{https://www-mars.lmd.jussieu.fr/}.\par
The MCD provided variables are a) atmospheric density in $\mathrm{kg/m}^3$, b) horizontal wind speed in m/s, and c) incident solar flux on a horizontal surface in $\mathrm{W/m}^2$. All variables were taken for an altitude of $50\,\mathrm{m}$ above the local surface (corresponding to the wind turbine's hub height) with a resolution of $5.625\degree \times 3.75\degree$ longitude-latitude. Additionally, the data includes full-day hourly observations during the solstices and equinoxes, i.e, the northern vernal equinox (VE) at solar longitude $(\text{Ls})=0\degree$, northern summer solstice (SS) at $\text{Ls}=90\degree$, northern autumnal equinox (AE) at $\text{Ls}=180\degree$, and northern winter solstice (WS) at $\text{Ls}=270\degree$.\par

\subsection{Solar Power on Mars}\label{RefSAW}
The solar flux reaching the Martian surface is less than half of that reaching Earth \cite{appelbaum1990solar}. Solar flux on Mars varies widely with latitude due to the eccentricity of Mars' orbit. Furthermore, global and local dust storms reduce the overall solar flux received at the surface. Another factor is dust accumulation on solar panels, which can block up to $0.14\%$ of the area per Martian day \cite{elliott2023mars}.\par
Recently, NASA proposed the Solar Arrays With Storage (SAWS) module \cite{elliott2023mars}. Each module, contained within a lander with a diameter of 10 meters, can deploy a $1\,000\,\mathrm{m}^2$ solar array that is expected to continuously generate $10\,\mathrm{kW}$.\par 
The power produced by the PV panels can be calculated as
\begin{equation}
    \text{Power}_\text{PV}=\phi \eta A,
\end{equation}
where $\phi$ is the solar flux incident on a horizontal surface in $\mathrm{W/m}^2$, $\eta$ is the efficiency factor, and $A$ is the total area of the PV array in $\mathrm{m}^2$. This study assumes that $\eta$ and $A$ are constant, while $\phi$ is time-varying and provided by the MCD. The efficiency of triple-junction space-grade solar cells is approximately $29\%$ \cite{elliott2023mars}.\par

\begin{figure}[!t]
\centering
\includegraphics[width=1\columnwidth]{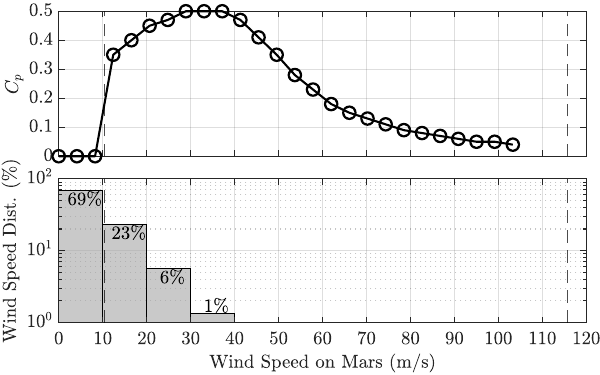}
\caption{Upper panel: efficiency factor ($C_p$) versus wind speed on Mars for the Enercon E33 (for $\rho=0.017\,\mathrm{kg/m}^3$). Lower panel: Global wind speed distribution percentage on Mars for all seasons including global dust storms. The dashed lines represent the cut-in and cut-out speeds.}
\label{figCp}
\end{figure}

\subsection{Wind Power on Mars}
Due to the low density of the Martian Atmosphere, the force produced by Martian winds is approximately $99\%$ less than that of Earth \cite{hartwick2023assessment}. Therefore, it is important to design a low Reynolds number turbine that can utilize the characteristics of Martian winds.\par
The power produced by a wind turbine can be represented by \cite{hartwick2023assessment}
\begin{equation}
    \text{Power}_\text{wind}=\frac{1}{2}\rho v^3 C_p A,
\end{equation}
where $\rho$ is the air density in $\mathrm{kg/m}^3$, $v$ is the wind speed at blade height in $\mathrm{m/s}$, $C_p$ is the efficiency factor which is bounded by an upper limit of $\approx 0.593$ (Betz limit), and $A$ is the rotor swept area in $\mathrm{m}^2$. This work assumes that $\rho$ and $v$ are time-varying and provided by the MCD, while $A$ is constant. Additionally, the $C_p$ values provided by the wind turbine's manufacturer were scaled to Mars's air density and wind speeds (see upper panel of Fig. \ref{figCp} and (3) in \cite{hartwick2023assessment} for more details).\par
It is noted that, for the E33, assuming $\rho_\text{Earth}=1.225\,\mathrm{kg/m}^3$ and $\rho_\text{Mars}=0.017\,\mathrm{kg/m}^3$, the cut-in speed on Mars is $10.3\,\mathrm{m/s}$ and the cut-out speed is $115.7\,\mathrm{m/s}$. These values are considerably higher than the cut-in and cut-out wind speeds of the E33 on Earth at $2.5\,\mathrm{m/s}$ and $28\,\mathrm{m/s}$, respectively. This is due to the difference in $\rho$ between the two planets. Resultantly, only $30\%$ of the global Martian wind speed distribution exceeds the cut-in speed (see lower panel of Fig. \ref{figCp}).\par

\subsection{BESS}
To compensate for the lack of nighttime solar flux, and to counter the intermittency of wind energy, a BESS must be incorporated as a backup to PV and wind systems. The sizing of the BESS depends on the weight-to-cost ratio as it must be transported from Earth to Mars. For instance, in \cite{abel2022photovoltaics} the proposed energy storage systems were sized to support a full day ($312\,\mathrm{kWh}$).\par
The SAWS module \cite{elliott2023mars} uses regenerative fuel cells (RFC) to support nighttime loads. Some of the benefits of RFC compared to traditional lithium-ion batteries are their faster recharge time, and resilience to fluctuations in depth of discharge \cite{arsalis2022comparative}.\par

\subsection{Martian Dust Storms}
Dust storms that occur on the Martian surface can be grouped into two categories: (1) global dust storms and (2) local dust storms. Global dust storms cover large portions of the Martian surface and may last $35$--$70$ days or more. In contrast, local dust storms are less intense, cover smaller regions, and last for a few days \cite{appelbaum1990solar}.\par
The effect of both global and local dust storms on the performance of any proposed power system must be evaluated using a climate model such as the MCD. This ensures the proposed system can withstand the harsh realities of the Martian environment and provides a continuous power supply to critical parts of the base such as life support modules.\par 
This study utilizes two scenarios provided by the MCD, the \emph{Climatology---ave solar} scenario, representing standard Martian conditions (including local dust storms), and the \emph{Dust Storm---ave solar} scenario, representing a global dust storm period. Since global dust storms typically occur during the northern AE--WS period \cite{appelbaum1990solar}, they are represented in this study by the AE-GS and WS-GS notations.\par

\section{Analysis \& Discussion}\label{Sec3}

\begin{figure}[htbp]
\centering
\includegraphics[width=1\columnwidth]{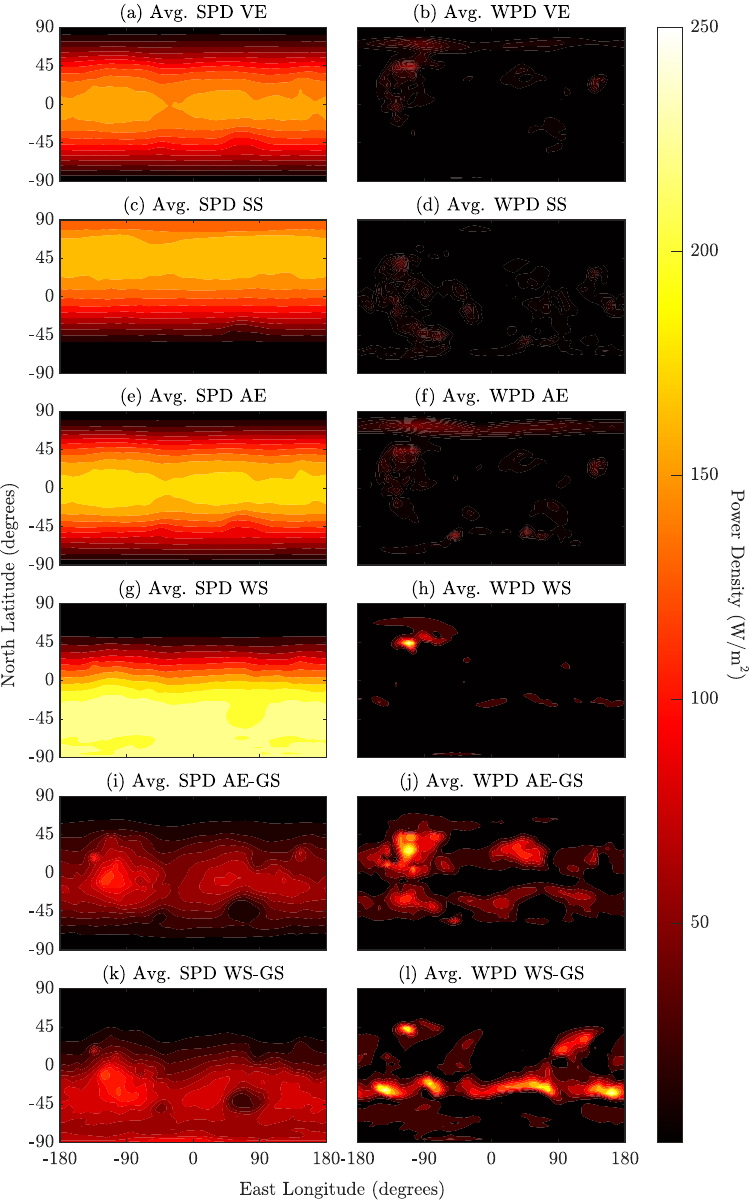}
\caption{Daily average Solar Power Density (first column) and Wind Power Density (second column) in $\mathrm{W/m}^2$ per season. Note that GS represents a global dust storm period.}
\label{fig1}
\end{figure}

To understand the diurnal and seasonal variations in solar and wind power, the daily average solar power density (SPD) and wind power density (WPD) over the entire Martian surface are illustrated in Fig. \ref{fig1}. Note that $\text{SPD}=\phi$ and $\text{WPD}=\frac{1}{2}\rho v^3$ \cite{hartwick2023assessment}.\par
While SPD values vary considerably due to different seasonal conditions, the equatorial regions consistently have higher average SPD than other regions. Additionally, SPD in the southern hemisphere during the northern winter solstice is higher than during other seasons due to the eccentricity of the Martian orbit \cite{levine1977solar}. A significant reduction of SPD during global dust storms can be seen in Fig. \ref{fig1}(i,k) compared to the same seasons without global dust storms in Fig. \ref{fig1}(e,g).\par
The spatial and temporal variability of WPD is illustrated in Fig. \ref{fig1}, revealing unique seasonal patterns. For instance, the western hemisphere consistently shows higher average WPD than the eastern hemisphere. Furthermore, WPD near the northern pole appears to be somewhat higher than near the southern pole during the equinoxes. Yet, the most striking feature of Fig. \ref{fig1} is the significant increase of WPD during global dust storms (Fig. \ref{fig1}(j,l)), compared to the same seasons without global dust storms (Fig. \ref{fig1}(f,h)).\par
Fig. \ref{fig1}(i--l) shows that the reduction of solar flux during global dust storms coincides with the increase in wind speed. This indicates that wind energy can effectively complement solar energy during global dust storms \cite{hartwick2023assessment}.\par
\break
Throughout this study, we assume the following: 
\begin{itemize}
    \item The hourly load is assumed to be $13\,\mathrm{kW}$ (see Section \ref{RefEn}).
    \item A single PV array is equivalent to a single SAWS module (see Section \ref{RefSAW}) covering an area of $1\,000\,\mathrm{m}^2$ with an efficiency rating of $29\%$.
    \item This study utilizes the Enercon E33 wind turbine with a hub height of $50\,\mathrm{m}$ and a rotor diameter of $33.4\,\mathrm{m}$ to remain consistent with related work \cite{hartwick2023assessment}.
    \item The BESS, assumed to be empty at the beginning of the day, is charged to its full capacity of $312\,\mathrm{kWh}$ using generated power that exceeds the load demand.
    \item The day is assumed to begin at midday ($12{:}00$ Martian time) and end after $24$ hours to allow ample time for battery charging.
    \item Similar to \cite{hartwick2023assessment}, the effect of dust accumulation on PV panel efficiency is not considered in this study.
\end{itemize}%

\begin{figure}[t!]
\centering
\includegraphics[width=1\columnwidth]{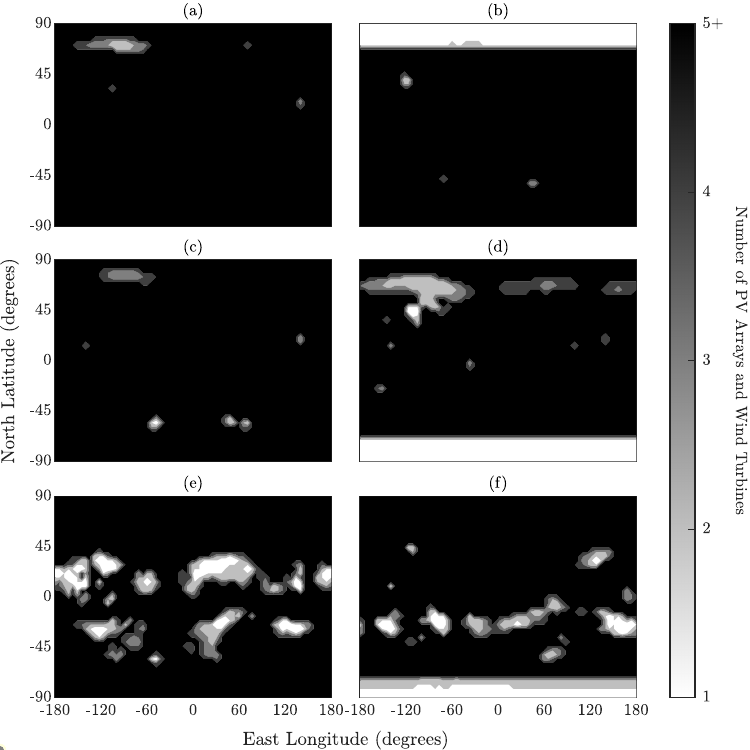}
\caption{The number of PV arrays and wind turbines required to power the PV-wind system for an entire day during: a) vernal equinox, b) summer solstice, c) autumnal equinox, d) winter solstice, e) autumnal equinox during a global dust storm, and f) winter solstice during a global dust storm.}
\label{fig4}
\end{figure}

\begin{figure}[t!]
\centering
\includegraphics[width=1\columnwidth]{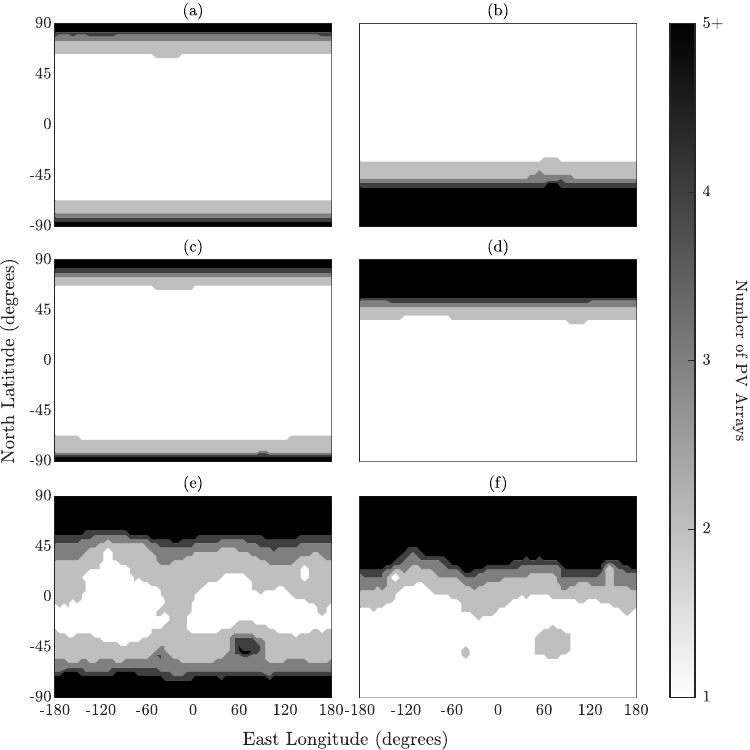}
\caption{The number of PV arrays required to power the PV-BESS system for an entire day during a) vernal equinox, b) summer solstice, c) autumnal equinox, d) winter solstice, e) autumnal equinox during a global dust storm, and f) winter solstice during a global dust storm.}
\label{fig5}
\end{figure}

\begin{figure}[t!]
\centering
\includegraphics[width=1\columnwidth]{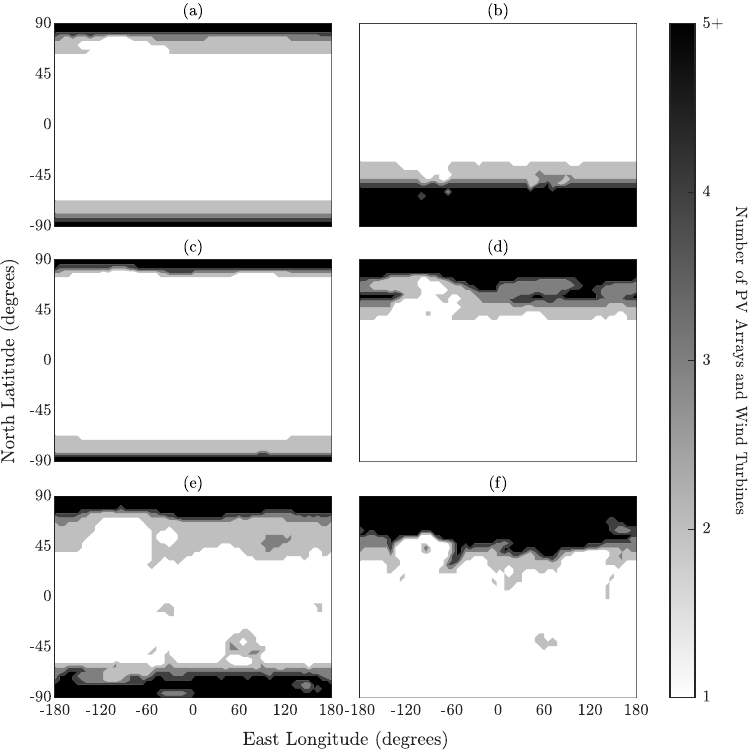}
\caption{The number of PV arrays and wind turbines required to power the PV-wind-BESS system for an entire day during: a) vernal equinox, b) summer solstice, c) autumnal equinox, d) winter solstice, e) autumnal equinox during a global dust storm, and f) winter solstice during a global dust storm.}
\label{fig7}
\end{figure}

\subsection{Global Analysis}

Figs. \ref{fig4}--\ref{fig7} present the global seasonal analysis of the PV-wind, PV-BESS, and PV-wind-BESS configurations, respectively, in terms of the number of PV arrays and wind turbines (from $1$ to $5+$) required to meet the demand of a Mars base for a full day. This analysis assumes a $1{:}1$ ratio of PV arrays to wind turbines.\par

Fig. \ref{fig4} shows the number of PV arrays and wind turbines, in a PV-wind power system, needed to support the Mars base for a full day during the equinoxes and solstices. Due to the absence of a BESS in this configuration, only the northern and southern poles during the northern summer and winter solstices, respectively, and some higher latitude regions in the northwest can be supported for a full day. During global dust storms, some equatorial and low-latitude regions can also be supported by the system. This is due to the elevated nighttime power production of the wind turbines, which compensates for the lack of solar energy.\par

Fig. \ref{fig5} shows the number of PV arrays, in a PV-BESS power system, needed to support the Mars base for a full day during the equinoxes and solstices. Fig. \ref{fig5}(a, c) indicates that a single PV array with a BESS can support the majority of the Martian surface during the equinoxes. However, this does not apply to the northern summer and winter solstices, where large portions of the southern and northern hemispheres do not receive sufficient solar flux to support the base. 
Furthermore, a single PV array with a BESS is insufficient to support the base during global dust storms unless multiple redundant PV arrays are installed, which is not cost-effective.\par

Fig. \ref{fig7} shows the number of PV arrays and wind turbines, in a PV-wind-BESS power system, needed to support the Mars base for a full day during the equinoxes and solstices. Similar to Fig. \ref{fig5}, Fig. \ref{fig7}(a, c) shows that the majority of the Martian surface can be supported by a single PV, wind turbine, and a BESS during the equinoxes. Moreover, the inclusion of the wind turbine extends the system's applicability to a wider area of the Martian surface during the solstices and global dust storms. Therefore, a sustainable Mars base utilizing renewable energy must include a PV array, a wind turbine, and BESS to ensure uninterrupted power supply across different seasonal and climatic conditions. The PV array provides the majority of the energy, while the wind turbine acts as a secondary/backup power source during periods of limited solar flux such as global dust storms, and the BESS stores any excess power to be used during outages.\par

\begin{figure}[t!]
\centering
\includegraphics[width=1\columnwidth]{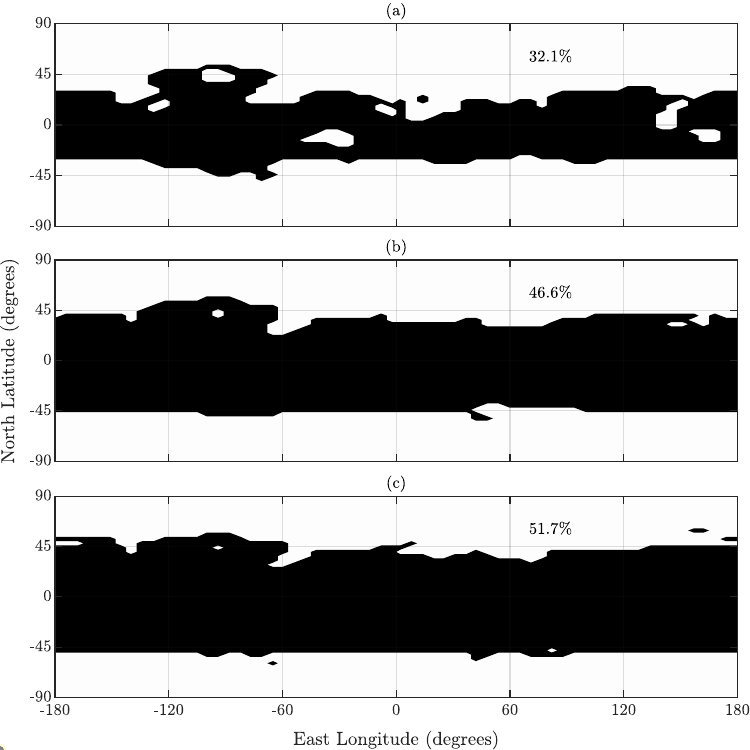}
\caption{Areas that can be supported by a) 1, b) 2, and c) 3 PV arrays and wind turbines (with BESS) through all seasons, and global dust storms. The percentages at the top right represent the portion of the Martian surface that each system can support.}
\label{fig8}
\end{figure}

Fig. \ref{fig8} compares the areas that can be supported by 1, 2, and 3 PV arrays and wind turbines with BESS throughout the equinoxes and solstices, including global dust storm periods. It is shown that a single PV array, wind turbine, and BESS can support a base located on $32.1\%$ of the Martian surface. On the other hand, using two sets of PV arrays and wind turbines with a BESS can support a base located on $46.6\%$ of the Martian surface, and $51.7\%$ for three sets. The majority of the supported area falls within the equatorial/low-latitude regions. Additionally, using three sets of PV arrays and wind turbines can support a base located within the majority of the $45\degree$ to $-45\degree$ latitude region.\par

\subsection{Site Analysis}

The ``First Landing Site/Exploration Zone Workshop for Human Missions to the Surface of Mars (FLSW),'' organized by NASA in 2015, aimed to identify and discuss candidate sites for human landing, habitation, and work on the Martian surface \cite{bussey2016human}. A total of 47 proposed sites were presented at the workshop. These sites were selected based on factors such as terrain stability, and their proximity to potential water sources and regions of scientific interest. While the wind and solar energy potentials of these sites were studied in \cite{hartwick2023assessment}, this study evaluates the seasonal performance of a PV-wind-battery power system for these sites.\par

\begin{table*}[htbp]
\caption{Seasonal surplus energy production of different sites (Note: minimum surplus energy per site is underlined)}
\begin{center}
\begin{tabular}{|c|cccccc|}
\hline
\multirow{2}{*}{\textbf{\begin{tabular}[c]{@{}c@{}}Site Name [Ref.]\\ (Latitude, Longitude)\end{tabular}}} & \multicolumn{6}{c|}{\textbf{\begin{tabular}[c]{@{}c@{}}Surplus Energy Produced\\ (Percentage of total energy produced by PV array, by wind turbine)\end{tabular}}} \\ \cline{2-7} 
 & \multicolumn{1}{c|}{\textbf{VE}} & \multicolumn{1}{c|}{\textbf{SS}} & \multicolumn{1}{c|}{\textbf{AE}} & \multicolumn{1}{c|}{\textbf{WS}} & \multicolumn{1}{c|}{\textbf{AE-GS}} & \textbf{WS-GS} \\ \hline

\begin{tabular}[c]{@{}c@{}}Acidalia Planitia \cite{kochemasov2015vallis}\\ ($20.00\degree$N, $-40.00\degree$E) \end{tabular} & \multicolumn{1}{c|}{\begin{tabular}[c]{@{}c@{}}$705\,\mathrm{kWh}$\\ $(97.1\%,2.9\%)$\end{tabular}} & \multicolumn{1}{c|}{\begin{tabular}[c]{@{}c@{}}$796\,\mathrm{kWh}$\\ $(100\%,0\%)$\end{tabular}} & \multicolumn{1}{c|}{\begin{tabular}[c]{@{}c@{}}$868\,\mathrm{kWh}$\\ $(98.2\%,1.8\%)$\end{tabular}} & \multicolumn{1}{c|}{\begin{tabular}[c]{@{}c@{}}$573\,\mathrm{kWh}$\\ $(81.2\%,18.8\%)$\end{tabular}} & \multicolumn{1}{c|}{\begin{tabular}[c]{@{}c@{}}\underline{$288\,\mathrm{kWh}$}\\ $(41.2\%,58.8\%)$\end{tabular}} & \begin{tabular}[c]{@{}c@{}}$428\,\mathrm{kWh}$\\ $(15.0\%,85.0\%)$\end{tabular} \\ \hline

\begin{tabular}[c]{@{}c@{}}Apollinaris Sulci \cite{kerber2015human,rice2015manned}\\ ($-12.67\degree$N, $176.67\degree$E) \end{tabular} & \multicolumn{1}{c|}{\begin{tabular}[c]{@{}c@{}}$780\,\mathrm{kWh}$\\ $(99.2\%,0.8\%)$\end{tabular}} & \multicolumn{1}{c|}{\begin{tabular}[c]{@{}c@{}}$487\,\mathrm{kWh}$\\ $(95.4\%,4.6\%)$\end{tabular}} & \multicolumn{1}{c|}{\begin{tabular}[c]{@{}c@{}}$922\,\mathrm{kWh}$\\ $(100\%,0\%)$\end{tabular}} & \multicolumn{1}{c|}{\begin{tabular}[c]{@{}c@{}}$1\,076\,\mathrm{kWh}$\\ $(99.2\%,0.8\%)$\end{tabular}} & \multicolumn{1}{c|}{\begin{tabular}[c]{@{}c@{}}\underline{$143\,\mathrm{kWh}$}\\ $(91.0\%,9.0\%)$\end{tabular}} & \begin{tabular}[c]{@{}c@{}}$196\,\mathrm{kWh}$\\ $(74.3\%,25.7\%)$\end{tabular} \\ \hline

\begin{tabular}[c]{@{}c@{}}Aram Chaos \cite{sibille2015aram}\\ ($2.42\degree$N, $-20.03\degree$E) \end{tabular} & \multicolumn{1}{c|}{\begin{tabular}[c]{@{}c@{}}$759\,\mathrm{kWh}$\\ $(100\%,0\%)$\end{tabular}} & \multicolumn{1}{c|}{\begin{tabular}[c]{@{}c@{}}$642\,\mathrm{kWh}$\\ $(100\%,0\%)$\end{tabular}} & \multicolumn{1}{c|}{\begin{tabular}[c]{@{}c@{}}$923\,\mathrm{kWh}$\\ $(100\%,0\%)$\end{tabular}} & \multicolumn{1}{c|}{\begin{tabular}[c]{@{}c@{}}$798\,\mathrm{kWh}$\\ $(98.5\%,1.5\%)$\end{tabular}} & \multicolumn{1}{c|}{\begin{tabular}[c]{@{}c@{}}$178\,\mathrm{kWh}$\\ $(73.0\%,27.0\%)$\end{tabular}} & \begin{tabular}[c]{@{}c@{}}\underline{$112\,\mathrm{kWh}$}\\ $(58.4\%,41.6\%)$\end{tabular} \\ \hline

\begin{tabular}[c]{@{}c@{}}Ausonia Cavus \cite{hamilton2015ausonia}\\ ($-32.00\degree$N, $96.50\degree$E) \end{tabular} & \multicolumn{1}{c|}{\begin{tabular}[c]{@{}c@{}}$648\,\mathrm{kWh}$\\ $(95.9\%,4.1\%)$\end{tabular}} & \multicolumn{1}{c|}{\begin{tabular}[c]{@{}c@{}}\underline{$140\,\mathrm{kWh}$}\\ $(92.3\%,7.7\%)$\end{tabular}} & \multicolumn{1}{c|}{\begin{tabular}[c]{@{}c@{}}$805\,\mathrm{kWh}$\\ $(92.7\%,7.3\%)$\end{tabular}} & \multicolumn{1}{c|}{\begin{tabular}[c]{@{}c@{}}$1\,323\,\mathrm{kWh}$\\ $(92.6\%,7.4\%)$\end{tabular}} & \multicolumn{1}{c|}{\begin{tabular}[c]{@{}c@{}}$522\,\mathrm{kWh}$\\ $(36.0\%,64.0\%)$\end{tabular}} & \begin{tabular}[c]{@{}c@{}}$544\,\mathrm{kWh}$\\ $(48.2\%,51.8\%)$\end{tabular} \\ \hline

\begin{tabular}[c]{@{}c@{}}Cerberus Fossae \cite{wright2015exploration}\\ ($10.00\degree$N, $162.00\degree$E) \end{tabular} & \multicolumn{1}{c|}{\begin{tabular}[c]{@{}c@{}}$751\,\mathrm{kWh}$\\ $(100\%,0\%)$\end{tabular}} & \multicolumn{1}{c|}{\begin{tabular}[c]{@{}c@{}}$720\,\mathrm{kWh}$\\ $(100\%,0\%)$\end{tabular}} & \multicolumn{1}{c|}{\begin{tabular}[c]{@{}c@{}}$916\,\mathrm{kWh}$\\ $(100\%,0\%)$\end{tabular}} & \multicolumn{1}{c|}{\begin{tabular}[c]{@{}c@{}}$651\,\mathrm{kWh}$\\ $(100\%,0\%)$\end{tabular}} & \multicolumn{1}{c|}{\begin{tabular}[c]{@{}c@{}}$231\,\mathrm{kWh}$\\ $(63.6\%,36.4\%)$\end{tabular}} & \begin{tabular}[c]{@{}c@{}}\underline{$197\,\mathrm{kWh}$}\\ $(38.8\%,61.2\%)$\end{tabular} \\ \hline

\begin{tabular}[c]{@{}c@{}}Chryse Planitia \cite{farrell2015mars}\\ ($22.30\degree$N, $-48.30\degree$E) \end{tabular} & \multicolumn{1}{c|}{\begin{tabular}[c]{@{}c@{}}$693\,\mathrm{kWh}$\\ $(97.7\%,2.3\%)$\end{tabular}} & \multicolumn{1}{c|}{\begin{tabular}[c]{@{}c@{}}$813\,\mathrm{kWh}$\\ $(100\%,0\%)$\end{tabular}} & \multicolumn{1}{c|}{\begin{tabular}[c]{@{}c@{}}$849\,\mathrm{kWh}$\\ $(98.9\%,1.1\%)$\end{tabular}} & \multicolumn{1}{c|}{\begin{tabular}[c]{@{}c@{}}$401\,\mathrm{kWh}$\\ $(96.0\%,4.0\%)$\end{tabular}} & \multicolumn{1}{c|}{\begin{tabular}[c]{@{}c@{}}$213\,\mathrm{kWh}$\\ $(47.2\%,52.8\%)$\end{tabular}} & \begin{tabular}[c]{@{}c@{}}\underline{$127\,\mathrm{kWh}$}\\ $(24.4\%,75.6\%)$\end{tabular} \\ \hline

\begin{tabular}[c]{@{}c@{}}Columbus Crater \cite{lynch2015exploring}\\ ($-29.00\degree$N, $-166.00\degree$E) \end{tabular} & \multicolumn{1}{c|}{\begin{tabular}[c]{@{}c@{}}$676\,\mathrm{kWh}$\\ $(100\%,0\%)$\end{tabular}} & \multicolumn{1}{c|}{\begin{tabular}[c]{@{}c@{}}\underline{$186\,\mathrm{kWh}$}\\ $(100\%,0\%)$\end{tabular}} & \multicolumn{1}{c|}{\begin{tabular}[c]{@{}c@{}}$799\,\mathrm{kWh}$\\ $(98.9\%,1.1\%)$\end{tabular}} & \multicolumn{1}{c|}{\begin{tabular}[c]{@{}c@{}}$1\,378\,\mathrm{kWh}$\\ $(93.6\%,6.4\%)$\end{tabular}} & \multicolumn{1}{c|}{\begin{tabular}[c]{@{}c@{}}$785\,\mathrm{kWh}$\\ $(40.6\%,59.4\%)$\end{tabular}} & \begin{tabular}[c]{@{}c@{}}$1411\,\mathrm{kWh}$\\ $(34.2\%,65.8\%)$\end{tabular} \\ \hline

\begin{tabular}[c]{@{}c@{}}Coprates Chasma \cite{mojarro2015human}\\ ($-11.68\degree$N, $-66.63\degree$E) \end{tabular} & \multicolumn{1}{c|}{\begin{tabular}[c]{@{}c@{}}$775\,\mathrm{kWh}$\\ $(100\%,0\%)$\end{tabular}} & \multicolumn{1}{c|}{\begin{tabular}[c]{@{}c@{}}$460\,\mathrm{kWh}$\\ $(100\%,0\%)$\end{tabular}} & \multicolumn{1}{c|}{\begin{tabular}[c]{@{}c@{}}$920\,\mathrm{kWh}$\\ $(100\%,0\%)$\end{tabular}} & \multicolumn{1}{c|}{\begin{tabular}[c]{@{}c@{}}$1\,075\,\mathrm{kWh}$\\ $(99.6\%,0.4\%)$\end{tabular}} & \multicolumn{1}{c|}{\begin{tabular}[c]{@{}c@{}}\underline{$228\,\mathrm{kWh}$}\\ $(86.2\%,13.8\%)$\end{tabular}} & \begin{tabular}[c]{@{}c@{}}$258\,\mathrm{kWh}$\\ $(76.8\%,23.2\%)$\end{tabular} \\ \hline

\begin{tabular}[c]{@{}c@{}}Firsoff Crater \cite{ori2015exploration}\\ ($0.39\degree$N, $-8.36\degree$E) \end{tabular} & \multicolumn{1}{c|}{\begin{tabular}[c]{@{}c@{}}$775\,\mathrm{kWh}$\\ $(100\%,0\%)$\end{tabular}} & \multicolumn{1}{c|}{\begin{tabular}[c]{@{}c@{}}$621\,\mathrm{kWh}$\\ $(100\%,0\%)$\end{tabular}} & \multicolumn{1}{c|}{\begin{tabular}[c]{@{}c@{}}$942\,\mathrm{kWh}$\\ $(100\%,0\%)$\end{tabular}} & \multicolumn{1}{c|}{\begin{tabular}[c]{@{}c@{}}$869\,\mathrm{kWh}$\\ $(96.8\%,3.2\%)$\end{tabular}} & \multicolumn{1}{c|}{\begin{tabular}[c]{@{}c@{}}$222\,\mathrm{kWh}$\\ $(72.1\%,27.9\%)$\end{tabular}} & \begin{tabular}[c]{@{}c@{}}\underline{$108\,\mathrm{kWh}$}\\ $(65.7\%,34.3\%)$\end{tabular} \\ \hline

\begin{tabular}[c]{@{}c@{}}Gale Crater \cite{calef2015assessing,montano2015ground,yun2015nasa}\\ ($-4.60\degree$N, $137.40\degree$E) \end{tabular} & \multicolumn{1}{c|}{\begin{tabular}[c]{@{}c@{}}$807\,\mathrm{kWh}$\\ $(97.9\%,2.1\%)$\end{tabular}} & \multicolumn{1}{c|}{\begin{tabular}[c]{@{}c@{}}$598\,\mathrm{kWh}$\\ $(95.0\%,5.0\%)$\end{tabular}} & \multicolumn{1}{c|}{\begin{tabular}[c]{@{}c@{}}$959\,\mathrm{kWh}$\\ $(98.2\%,1.8\%)$\end{tabular}} & \multicolumn{1}{c|}{\begin{tabular}[c]{@{}c@{}}$946\,\mathrm{kWh}$\\ $(99.1\%,0.9\%)$\end{tabular}} & \multicolumn{1}{c|}{\begin{tabular}[c]{@{}c@{}}$195\,\mathrm{kWh}$\\ $(80.2\%,19.8\%)$\end{tabular}} & \begin{tabular}[c]{@{}c@{}}\underline{$76\,\mathrm{kWh}$}\\ $(81.3\%,18.7\%)$\end{tabular} \\ \hline

\begin{tabular}[c]{@{}c@{}}Gusev Crater \cite{longo2015landing}\\ ($-14.50\degree$N, $175.40\degree$E) \end{tabular} & \multicolumn{1}{c|}{\begin{tabular}[c]{@{}c@{}}$783\,\mathrm{kWh}$\\ $(98.1\%,1.9\%)$\end{tabular}} & \multicolumn{1}{c|}{\begin{tabular}[c]{@{}c@{}}$479\,\mathrm{kWh}$\\ $(92.8\%,7.2\%)$\end{tabular}} & \multicolumn{1}{c|}{\begin{tabular}[c]{@{}c@{}}$938\,\mathrm{kWh}$\\ $(97.9\%,2.1\%)$\end{tabular}} & \multicolumn{1}{c|}{\begin{tabular}[c]{@{}c@{}}$1\,106\,\mathrm{kWh}$\\ $(98.9\%,1.1\%)$\end{tabular}} & \multicolumn{1}{c|}{\begin{tabular}[c]{@{}c@{}}\underline{$144\,\mathrm{kWh}$}\\ $(91.6\%,8.4\%)$\end{tabular}} & \begin{tabular}[c]{@{}c@{}}$220\,\mathrm{kWh}$\\ $(74.2\%,25.8\%)$\end{tabular} \\ \hline

\begin{tabular}[c]{@{}c@{}}Hadriacus Palus \cite{skinner2015considerations}\\ ($-26.84\degree$N, $77.47\degree$E) \end{tabular} & \multicolumn{1}{c|}{\begin{tabular}[c]{@{}c@{}}$706\,\mathrm{kWh}$\\ $(94.7\%,5.3\%)$\end{tabular}} & \multicolumn{1}{c|}{\begin{tabular}[c]{@{}c@{}}\underline{$194\,\mathrm{kWh}$}\\ $(97.7\%,2.3\%)$\end{tabular}} & \multicolumn{1}{c|}{\begin{tabular}[c]{@{}c@{}}$854\,\mathrm{kWh}$\\ $(93.3\%,6.7\%)$\end{tabular}} & \multicolumn{1}{c|}{\begin{tabular}[c]{@{}c@{}}$1\,498\,\mathrm{kWh}$\\ $(82.7\%,17.3\%)$\end{tabular}} & \multicolumn{1}{c|}{\begin{tabular}[c]{@{}c@{}}$366\,\mathrm{kWh}$\\ $(45.3\%,54.7\%)$\end{tabular}} & \begin{tabular}[c]{@{}c@{}}$2\,976\,\mathrm{kWh}$\\ $(12.3\%,87.7\%)$\end{tabular} \\ \hline

\begin{tabular}[c]{@{}c@{}}Hebrus Valles \cite{davila2015hebrus}\\ ($20.08\degree$N, $126.63\degree$E) \end{tabular} & \multicolumn{1}{c|}{\begin{tabular}[c]{@{}c@{}}$717\,\mathrm{kWh}$\\ $(97.4\%,2.6\%)$\end{tabular}} & \multicolumn{1}{c|}{\begin{tabular}[c]{@{}c@{}}$807\,\mathrm{kWh}$\\ $(97.8\%,2.2\%)$\end{tabular}} & \multicolumn{1}{c|}{\begin{tabular}[c]{@{}c@{}}$866\,\mathrm{kWh}$\\ $(98.6\%,1.4\%)$\end{tabular}} & \multicolumn{1}{c|}{\begin{tabular}[c]{@{}c@{}}$606\,\mathrm{kWh}$\\ $(79.1\%,20.9\%)$\end{tabular}} & \multicolumn{1}{c|}{\begin{tabular}[c]{@{}c@{}}\underline{$560\,\mathrm{kWh}$}\\ $(30.0\%,70.0\%)$\end{tabular}} & \begin{tabular}[c]{@{}c@{}}$742\,\mathrm{kWh}$\\ $(10.8\%,89.2\%)$\end{tabular} \\ \hline

\begin{tabular}[c]{@{}c@{}}Huygens Crater \cite{ackiss2015huygens}\\ ($-13.50\degree$N, $55.50\degree$E) \end{tabular} & \multicolumn{1}{c|}{\begin{tabular}[c]{@{}c@{}}$790\,\mathrm{kWh}$\\ $(99.4\%,0.6\%)$\end{tabular}} & \multicolumn{1}{c|}{\begin{tabular}[c]{@{}c@{}}\underline{$484\,\mathrm{kWh}$}\\ $(94.1\%,5.9\%)$\end{tabular}} & \multicolumn{1}{c|}{\begin{tabular}[c]{@{}c@{}}$952\,\mathrm{kWh}$\\ $(99.1\%,0.9\%)$\end{tabular}} & \multicolumn{1}{c|}{\begin{tabular}[c]{@{}c@{}}$1\,227\,\mathrm{kWh}$\\ $(92.5\%,7.5\%)$\end{tabular}} & \multicolumn{1}{c|}{\begin{tabular}[c]{@{}c@{}}$516\,\mathrm{kWh}$\\ $(61.2\%,38.8\%)$\end{tabular}} & \begin{tabular}[c]{@{}c@{}}$1\,257\,\mathrm{kWh}$\\ $(32.8\%,67.2\%)$\end{tabular} \\ \hline

\begin{tabular}[c]{@{}c@{}}Hypanis Delta \cite{gupta2015hypanis}\\ ($12.00\degree$N, $-45.50\degree$E) \end{tabular} & \multicolumn{1}{c|}{\begin{tabular}[c]{@{}c@{}}$769\,\mathrm{kWh}$\\ $(96.0\%,4.0\%)$\end{tabular}} & \multicolumn{1}{c|}{\begin{tabular}[c]{@{}c@{}}$813\,\mathrm{kWh}$\\ $(93.3\%,6.7\%)$\end{tabular}} & \multicolumn{1}{c|}{\begin{tabular}[c]{@{}c@{}}$949\,\mathrm{kWh}$\\ $(95.3\%,4.7\%)$\end{tabular}} & \multicolumn{1}{c|}{\begin{tabular}[c]{@{}c@{}}$690\,\mathrm{kWh}$\\ $(89.6\%,10.4\%)$\end{tabular}} & \multicolumn{1}{c|}{\begin{tabular}[c]{@{}c@{}}\underline{$299\,\mathrm{kWh}$}\\ $(53.2\%,46.8\%)$\end{tabular}} & \begin{tabular}[c]{@{}c@{}}$514\,\mathrm{kWh}$\\ $(21.9\%,78.1\%)$\end{tabular} \\ \hline

\begin{tabular}[c]{@{}c@{}}Mawrth Vallis \cite{horgan2015habitable}\\ ($24.30\degree$N, $-19.20\degree$E) \end{tabular} & \multicolumn{1}{c|}{\begin{tabular}[c]{@{}c@{}}$696\,\mathrm{kWh}$\\ $(97.7\%,2.3\%)$\end{tabular}} & \multicolumn{1}{c|}{\begin{tabular}[c]{@{}c@{}}$824\,\mathrm{kWh}$\\ $(100\%,0\%)$\end{tabular}} & \multicolumn{1}{c|}{\begin{tabular}[c]{@{}c@{}}$862\,\mathrm{kWh}$\\ $(97.7\%,2.3\%)$\end{tabular}} & \multicolumn{1}{c|}{\begin{tabular}[c]{@{}c@{}}$396\,\mathrm{kWh}$\\ $(91.9\%,8.1\%)$\end{tabular}} & \multicolumn{1}{c|}{\begin{tabular}[c]{@{}c@{}}$263\,\mathrm{kWh}$\\ $(46.0\%,54.0\%)$\end{tabular}} & \begin{tabular}[c]{@{}c@{}}\underline{$46\,\mathrm{kWh}$}\\ $(31.1\%,68.9\%)$\end{tabular} \\ \hline

\begin{tabular}[c]{@{}c@{}}McLaughlin Crater \cite{michalski2015mclaughlin}\\ ($21.90\degree$N, $-22.40\degree$E) \end{tabular} & \multicolumn{1}{c|}{\begin{tabular}[c]{@{}c@{}}$714\,\mathrm{kWh}$\\ $(96.9\%,3.1\%)$\end{tabular}} & \multicolumn{1}{c|}{\begin{tabular}[c]{@{}c@{}}$806\,\mathrm{kWh}$\\ $(100\%,0\%)$\end{tabular}} & \multicolumn{1}{c|}{\begin{tabular}[c]{@{}c@{}}$886\,\mathrm{kWh}$\\ $(96.8\%,3.2\%)$\end{tabular}} & \multicolumn{1}{c|}{\begin{tabular}[c]{@{}c@{}}$491\,\mathrm{kWh}$\\ $(85.8\%,14.2\%)$\end{tabular}} & \multicolumn{1}{c|}{\begin{tabular}[c]{@{}c@{}}$277\,\mathrm{kWh}$\\ $(44.5\%,55.5\%)$\end{tabular}} & \begin{tabular}[c]{@{}c@{}}\underline{$108\,\mathrm{kWh}$}\\ $(27.6\%,72.4\%)$\end{tabular} \\ \hline

\begin{tabular}[c]{@{}c@{}}Melas Chasma \cite{mcewen2015landing}\\ ($-11.70\degree$N, $-70.00\degree$E) \end{tabular} & \multicolumn{1}{c|}{\begin{tabular}[c]{@{}c@{}}$765\,\mathrm{kWh}$\\ $(100\%,0\%)$\end{tabular}} & \multicolumn{1}{c|}{\begin{tabular}[c]{@{}c@{}}$452\,\mathrm{kWh}$\\ $(100\%,0\%)$\end{tabular}} & \multicolumn{1}{c|}{\begin{tabular}[c]{@{}c@{}}$909\,\mathrm{kWh}$\\ $(100\%,0\%)$\end{tabular}} & \multicolumn{1}{c|}{\begin{tabular}[c]{@{}c@{}}$1\,049\,\mathrm{kWh}$\\ $(100\%,0\%)$\end{tabular}} & \multicolumn{1}{c|}{\begin{tabular}[c]{@{}c@{}}$163\,\mathrm{kWh}$\\ $(91.5\%,8.5\%)$\end{tabular}} & \begin{tabular}[c]{@{}c@{}}\underline{$151\,\mathrm{kWh}$}\\ $(86.2\%,13.8\%)$\end{tabular} \\ \hline

\begin{tabular}[c]{@{}c@{}}Meridiani Planum \cite{clarke2015first,cohen2015land}\\ ($-3.17\degree$N, $-4.52\degree$E) \end{tabular} & \multicolumn{1}{c|}{\begin{tabular}[c]{@{}c@{}}$782\,\mathrm{kWh}$\\ $(100\%,0\%)$\end{tabular}} & \multicolumn{1}{c|}{\begin{tabular}[c]{@{}c@{}}$580\,\mathrm{kWh}$\\ $(100\%,0\%)$\end{tabular}} & \multicolumn{1}{c|}{\begin{tabular}[c]{@{}c@{}}$950\,\mathrm{kWh}$\\ $(100\%,0\%)$\end{tabular}} & \multicolumn{1}{c|}{\begin{tabular}[c]{@{}c@{}}$921\,\mathrm{kWh}$\\ $(98.4\%,1.6\%)$\end{tabular}} & \multicolumn{1}{c|}{\begin{tabular}[c]{@{}c@{}}$180\,\mathrm{kWh}$\\ $(80.9\%,19.0\%)$\end{tabular}} & \begin{tabular}[c]{@{}c@{}}\underline{$98\,\mathrm{kWh}$}\\ $(74.6\%,25.4\%)$\end{tabular} \\ \hline

\begin{tabular}[c]{@{}c@{}}Nectaris Fossae \cite{boatwright2015southern}\\ ($-28.88\degree$N, $-59.71\degree$E) \end{tabular} & \multicolumn{1}{c|}{\begin{tabular}[c]{@{}c@{}}$676\,\mathrm{kWh}$\\ $(100\%,0\%)$\end{tabular}} & \multicolumn{1}{c|}{\begin{tabular}[c]{@{}c@{}}\underline{$189\,\mathrm{kWh}$}\\ $(100\%,0\%)$\end{tabular}} & \multicolumn{1}{c|}{\begin{tabular}[c]{@{}c@{}}$802\,\mathrm{kWh}$\\ $(100\%,0\%)$\end{tabular}} & \multicolumn{1}{c|}{\begin{tabular}[c]{@{}c@{}}$1\,311\,\mathrm{kWh}$\\ $(97.5\%,2.5\%)$\end{tabular}} & \multicolumn{1}{c|}{\begin{tabular}[c]{@{}c@{}}$515\,\mathrm{kWh}$\\ $(53.5\%,46.5\%)$\end{tabular}} & \begin{tabular}[c]{@{}c@{}}$957\,\mathrm{kWh}$\\ $(46.7\%,53.3\%)$\end{tabular} \\ \hline

\begin{tabular}[c]{@{}c@{}}Noctis Labyrinthus \cite{lee2015noctis}\\ ($-6.49\degree$N, $-92.45\degree$E) \end{tabular} & \multicolumn{1}{c|}{\begin{tabular}[c]{@{}c@{}}$859\,\mathrm{kWh}$\\ $(98.4\%,1.6\%)$\end{tabular}} & \multicolumn{1}{c|}{\begin{tabular}[c]{@{}c@{}}$546\,\mathrm{kWh}$\\ $(98.3\%,1.7\%)$\end{tabular}} & \multicolumn{1}{c|}{\begin{tabular}[c]{@{}c@{}}$1\,008\,\mathrm{kWh}$\\ $(98.8\%,1.2\%)$\end{tabular}} & \multicolumn{1}{c|}{\begin{tabular}[c]{@{}c@{}}$1\,080\,\mathrm{kWh}$\\ $(98.0\%,2.0\%)$\end{tabular}} & \multicolumn{1}{c|}{\begin{tabular}[c]{@{}c@{}}$644\,\mathrm{kWh}$\\ $(68.5\%,31.5\%)$\end{tabular}} & \begin{tabular}[c]{@{}c@{}}\underline{$414\,\mathrm{kWh}$}\\ $(80.5\%,19.5\%)$\end{tabular} \\ \hline

\begin{tabular}[c]{@{}c@{}}Sinus Meridiani \cite{wilkinson2015sinus}\\ ($-2.25\degree$N, $-6.43\degree$E) \end{tabular} & \multicolumn{1}{c|}{\begin{tabular}[c]{@{}c@{}}$779\,\mathrm{kWh}$\\ $(100\%,0\%)$\end{tabular}} & \multicolumn{1}{c|}{\begin{tabular}[c]{@{}c@{}}$591\,\mathrm{kWh}$\\ $(100\%,0\%)$\end{tabular}} & \multicolumn{1}{c|}{\begin{tabular}[c]{@{}c@{}}$946\,\mathrm{kWh}$\\ $(100\%,0\%)$\end{tabular}} & \multicolumn{1}{c|}{\begin{tabular}[c]{@{}c@{}}$906\,\mathrm{kWh}$\\ $(98.0\%,2.0\%)$\end{tabular}} & \multicolumn{1}{c|}{\begin{tabular}[c]{@{}c@{}}$196\,\mathrm{kWh}$\\ $(77.4\%,22.6\%)$\end{tabular}} & \begin{tabular}[c]{@{}c@{}}\underline{$104\,\mathrm{kWh}$}\\ $(71.4\%,28.6\%)$\end{tabular} \\ \hline

\begin{tabular}[c]{@{}c@{}}Valles Marineris \cite{mitchell2015equatorial}\\ ($-18.83\degree$N, $-49.21\degree$E) \end{tabular} & \multicolumn{1}{c|}{\begin{tabular}[c]{@{}c@{}}$736\,\mathrm{kWh}$\\ $(100\%,0\%)$\end{tabular}} & \multicolumn{1}{c|}{\begin{tabular}[c]{@{}c@{}}$365\,\mathrm{kWh}$\\ $(100\%,0\%)$\end{tabular}} & \multicolumn{1}{c|}{\begin{tabular}[c]{@{}c@{}}$878\,\mathrm{kWh}$\\ $(100\%,0\%)$\end{tabular}} & \multicolumn{1}{c|}{\begin{tabular}[c]{@{}c@{}}$1\,186\,\mathrm{kWh}$\\ $(99.0\%,1.0\%)$\end{tabular}} & \multicolumn{1}{c|}{\begin{tabular}[c]{@{}c@{}}\underline{$280\,\mathrm{kWh}$}\\ $(77.8\%,22.2\%)$\end{tabular}} & \begin{tabular}[c]{@{}c@{}}$337\,\mathrm{kWh}$\\ $(78.0\%,22.0\%)$\end{tabular} \\ \hline

\begin{tabular}[c]{@{}c@{}}Zephyria Planum \cite{yakovlev2015hills}\\ ($-1.37\degree$N, $157.13\degree$E) \end{tabular} & \multicolumn{1}{c|}{\begin{tabular}[c]{@{}c@{}}$778\,\mathrm{kWh}$\\ $(100\%,0\%)$\end{tabular}} & \multicolumn{1}{c|}{\begin{tabular}[c]{@{}c@{}}$640\,\mathrm{kWh}$\\ $(95.3\%,4.7\%)$\end{tabular}} & \multicolumn{1}{c|}{\begin{tabular}[c]{@{}c@{}}$939\,\mathrm{kWh}$\\ $(100\%,0\%)$\end{tabular}} & \multicolumn{1}{c|}{\begin{tabular}[c]{@{}c@{}}$867\,\mathrm{kWh}$\\ $(99.3\%,0.7\%)$\end{tabular}} & \multicolumn{1}{c|}{\begin{tabular}[c]{@{}c@{}}\underline{$149\,\mathrm{kWh}$}\\ $(79.2\%,20.8\%)$\end{tabular}} & \begin{tabular}[c]{@{}c@{}}$197\,\mathrm{kWh}$\\ $(51.1\%,48.9\%)$\end{tabular} \\ \hline

\end{tabular}\label{tab1}
\end{center}
\end{table*}

To determine if a single PV array, wind turbine, and BESS can support a base at the proposed $47$ sites, their coordinates were cross-referenced with the area highlighted in Fig. \ref{fig8}(a). A total of $24$ sites overlapped with this region, indicating that the load demand of a base at these sites can be supported throughout the equinoxes and solstices, even during global dust storms. These sites are listed in Table \ref{tab1}, which includes their seasonal performance in terms of the surplus energy produced by the power system. Additionally, Table \ref{tab1} includes the percentage of total energy produced by the PV array and wind turbine. In this study, surplus energy refers to the difference between the produced energy and the load. This surplus energy can be stored in the battery to mitigate unforeseen fluctuations in power production.\par

The percentage contribution of the wind turbine to the total power production during the solstices and equinoxes without global dust storms ranges from $0$--$20.9\%$, with the highest percentages typically occurring during the winter solstice. During global dust storms, this range increases to $8.4$--$89.2\%$. This shows that all sites benefit to some extent from wind-generated power during global dust storms.\par

The minimum seasonal surplus energy per site is underlined in Table \ref{tab1}. The minimum surplus energy was observed either during a global dust storm period or the summer solstice for sites in the southern hemisphere. The top three sites with the largest minimum surplus energy were Hebrus Valles with $560\,\mathrm{kWh}$ during AE-GS, Huygens Crater with $484\,\mathrm{kWh}$ during the summer solstice, and Noctis Labyrinthus with $414\,\mathrm{kWh}$ during WS-GS. These three sites are the most promising for future settlements from a power production perspective, as they produced at least $2.3$--$2.8\times$ the amount of energy required to run the base. On the other hand, the sites with the least minimum surplus energy were Mawrth Vallis, Gale Crater, and Meridiani Planum. All three sites had their minimum surplus energy during WS-GS.\par

The analysis above shows that a Mars base located on any of the $24$ sites listed in Table \ref{tab1}, with a $13\,\mathrm{kW}$ load, can be continuously supplied by a single PV array, wind turbine, and BESS throughout the solstices and equinoxes, including global dust storm periods. For larger loads, i.e., bigger bases, a modular design can be adopted where each $13\,\mathrm{kW}$ module is supplied by one $1\,000\,\mathrm{m}^2$ PV array, a $33.4\,\mathrm{m}$ diameter wind turbine, and a BESS with a full-day capacity ($312\,\mathrm{kWh}$).\par

\section{Conclusion}\label{Sec4}
This study demonstrates the viability of a PV-wind-battery power system for sustaining a Mars base under various seasonal and climatic conditions. Simulations using the Mars Climate Database revealed that the hybrid power system can effectively support the base despite challenges posed by seasonal changes, diurnal cycles, and dust storms. While the PV arrays serve as the primary energy source, the wind turbines provide essential backup during nighttime and dust storms. A single $1\,000\,\mathrm{m}^2$ PV array, coupled with a $33.4\,\mathrm{m}$ diameter wind turbine and a $312\,\mathrm{kWh}$ battery, can support a six-person Mars base across $32.1\%$ of the Martian surface during the equinoxes and solstices, including global dust storms. This coverage increases to $51.7\%$ with the deployment of three PV arrays and three wind turbines. Additionally, it is feasible to support a base at $24$ of the sites proposed in the ``First Landing Site/Exploration Zone Workshop for Human Missions to the Surface of Mars,'' even during global dust storms, using a single PV array, wind turbine, and battery. Among these, Hebrus Valles, Huygens Crater, and Noctis Labyrinthus were identified as the most promising in terms of energy production. These findings provide a foundational basis for future research into hybrid power systems for Mars exploration, highlighting their potential to significantly enhance the sustainability of human missions to the Martian surface.\par

% \section*{Acknowledgment}
% This work used the Mars Climate Database model, which is publicly available at \url{https://www-mars.lmd.jussieu.fr/}.

\section*{Acknowledgment}
We thank Francois Forget for his help with the MCD model.


\bibliographystyle{IEEEtran}
\bibliography{IEEEabrv,Manuscript.bib}
\end{document}